\RequirePackage{fix-cm}
\documentclass[twocolumn]{svjour3}          
\smartqed  
\usepackage{flafter}%
\usepackage{amsmath}
\usepackage{graphicx}
\usepackage{epstopdf}
\usepackage[utf8]{inputenc}
\usepackage{mathptmx}      
\usepackage[colorlinks=true, pdfborder={0 0 0}]{hyperref}
\usepackage{latexsym}

 \journalname{Mobile Networks and Applications}
\begin{document}

\title{Dwell Time Prediction Model for Minimizing Unnecessary Handovers in Heterogenous Wireless Networks, Considering Amoebic Shaped Coverage Region  
}

\titlerunning{Minimizing Unnecessary Handover in Heterogenous Wireless Networks}        

\author{Babatunji Omoniwa         \and
        Riaz Hussain 
}


\institute{B. Omoniwa \at
              COMSATS Institute of Information Technology, Park Road, Islamabad, Pakistan \& \\
              National Mathematical Centre, Abuja, Nigeria.\\
              \email{tunjiomoniwa@gmail.com}
           \and
          R. Hussain \at
          COMSATS Institute of Information Technology, Park Road, Islamabad, Pakistan \\
          Tel.: +92-33-35638665
          \email{rhussain@comsats.edu.pk}
}

\date{Received: date / Accepted: date}

\maketitle

\begin{abstract}
Over the years, vertical handover necessity estimation has attracted the interest of numerous researchers. Despite the attractive benefits of integrating different wireless platforms, mobile users are confronted with the issue of detrimental handover. This paper used extensive geometric and probability analysis in modelling the coverage area of a WLAN cell. Thus, presents a realistic and novel model with an attempt to minimize unnecessary handover and handover failure of a mobile node (MN) traversing the WLAN cell from a third generation (3G) network. The dwell time is estimated along with the threshold values to ensure an optimal handover decision by the MN, while the probability of unnecessary handover and handover failure are kept within tolerable bounds. Monte-Carlo simulations were carried out to show the behaviour of the proposed model. Results were validated by comparing this model with existing models for unnecessary handover minimization.
\keywords{Vertical handover necessity estimation \and Unnecessary handover \and Handover failure \and Monte-Carlo \and Dwell time\and WLAN \and 3G}
\end{abstract}

\section{Introduction}
\label{sec:1}
With rapid growth in the use of the Internet and wireless services, the challenge to support generalized mobility and provision of ubiquitous services to users while integrating diverse access technologies (GSM, 3G, 4G, WLAN, WiMAX and Bluetooth), has attracted research attention. Due to increased demand for mobile data, users now require access networks that use multiple layers (macro as well as micro cells), and multiple technologies to meet growing needs. As a mobile node (MN) moves within a heterogenous environment, satisfactory quality of service (QoS) is desired by ensuring efficient vertical handover.

Vertical handover can be defined as when an MN moves from one access network to another type of access network while maintaining the live call or session. In contrast to horizontal handovers, vertical handovers can be instigated for convenience rather than connectivity purposes~\cite{RefJ6}. As such, the choice to perform vertical handover may depend on factors such as available bandwidth, received signal strength (RSS), access cost, dwell time, security, speed, etc. \cite{RefJ1,RefJ2,RefJ3,RefJ4,RefJ5,RefJ6}. For optimal decision making, it is imperative to weigh the benefits against the detriments before initiating a vertical handover. Related works on vertical handover can be grouped into Three types \cite{RefJ5}, the RSS-based, Cost-based and Other related works which are briefly described.\begin{description}
                                    \item[\textit{\textbf{RSS-based related works}}:] Several RSS-based handover algorithms have been developed for wireless communications. A novel algorithm was developed using the concept of dynamic boundary area to support seamless vertical handover between the 3G and WLAN in \cite{RefJ4}. A traveling distance prediction based handover decision method \cite{RefJ2}, dwell time prediction model \cite{RefJ3} and a linear approximation of the travelled distance \cite{RefJ1} were proposed to minimize the probability of unnecessary handover. However, the geometric models considered were not of a realistic coverage cell shape.
                                    \item[\textit{\textbf{Cost-based related works}}:] Handover cost is a function of the available bandwidth, security, power consumption and the monetary cost \cite{RefJ11}. As the need for voice and video services rise, available bandwidth, power consumption, security, etc., will be the major factors to indicate network conditions and determination of triggering the handovers. In \cite{RefJ9}, available bandwidth and monetary cost were used as metrics for handover decisions. Cost-based algorithms are usually complex as they require collecting and normalization of different network metrics.
                                    \item[\textit{\textbf{Other related works}}:] An analytical framework to evaluate vertical handover algorithms with new extensions for traditional hysteresis based and dwell timer-based algorithms was proposed in \cite{RefJ10}. Using probability approach, \cite{RefJ7} worked on the assessment of a Wrong Decision Probability (WDP), which assures a trade-off between network performance maximization and mitigation of the ping-pong effect. The proposed algorithm was able to reduce the vertical handover frequency, thus minimizing the unnecessary handovers and maximizing the throughput by keeping the received bits as high as possible.
                                  \end{description}

The focus of this paper is to introduce an amoebic based geometric model that extends the ideal circular coverage model employed in previous works. The work considers the RSS-based dwell time approach. RSS-based algorithms are easy to implement, however, these algorithms are seriously limited by slow fading \cite{RefJ8}. Slow fading can be caused by events such as shadowing, where a large obstruction obscures the main signal path between the transmitter and the receiver. This paper presents a novel and realistic model that depicts the actual behaviour of a WLAN coverage area, considering the effects of slow fading. The proposed model will ensure an efficient minimization of the probability of unnecessary handover and handover failure for an MN traversing a realistic WLAN cell from a 3G network by the following factors:\begin{itemize}
          \item The WLAN cell shape is not exactly circular, but irregular.
          \item The cell shape changes with changes in nature of obstruction at different instances, humidity, temperature etc.
        \end{itemize}

The remainder of this paper is organized as follows: Section \ref{sec:2} describes the Amoebic Wireless Coverage Concept. Section \ref{sec:3} presents the Geometric and Probabilistic analysis of the proposed Amoebic-based dwell time model.
The Handover decision is specified in Section \ref{sec:4}. Section \ref{sec:5} evaluates the performance of the proposed model and compares the results with those of other approaches. Section \ref{sec:6} draws conclusions.

\section{Amoebic Wireless Coverage Concept}
\label{sec:2}
Both theoretical and empirical propagation models show that average received signal power decreases logarithmically with distance.\footnote{~$PL(d)_{dB}= P(transmitted)_{dB} - P(received)_{dB} = PL(d_0) + 10\beta\log(d/d_0) + R_{\sigma}$, \\Where ~$R_{\sigma}$ is the Gaussian random variable with standard deviation, ~$\sigma$ and ~$\beta$ is the path loss exponent.} There exist a number of factors, apart from the frequency and the distance that influence losses encountered by propagated signals from the AP to the MN. In order to accurately model a wireless coverage area the factors that must be considered are:\begin{itemize}
                         \item the height of the MN antenna;
                         \item the height of the AP relative to the surrounding terrain;
                         \item the terrain irregularity (undulation or roughness);
                         \item the land usage in the surroundings of MN: urban, suburban, rural, open, etc.
                       \end{itemize}

Due to these effects the coverage region does not remain circular, but of an irregular shape and this shape also changes with time --- thus we called it amoebic shaped coverage region. This paper presents an Amoebic WLAN cell which gives a realistic representation of the wireless coverage with a perspective of the shadowing concept. There are three different rates of variation as wireless signals are propagated away from the AP: we have (i) the very slow variation, called path loss, which is a function of distance between the AP and MN, (ii) slow variation, which results from shadowing effects, and (iii) fast variation, due to multipath. Signal variations caused by multipath, in the case where the direct signal is assumed to be totally blocked, are usually represented by a Rayleigh distribution\cite{RefB3}. The slow received signal variability due to shadowing is usually assumed to follow a Gaussian distribution~\cite{RefB3}.
\begin{figure*}
 \begin{center}
 \includegraphics[width=1\textwidth]{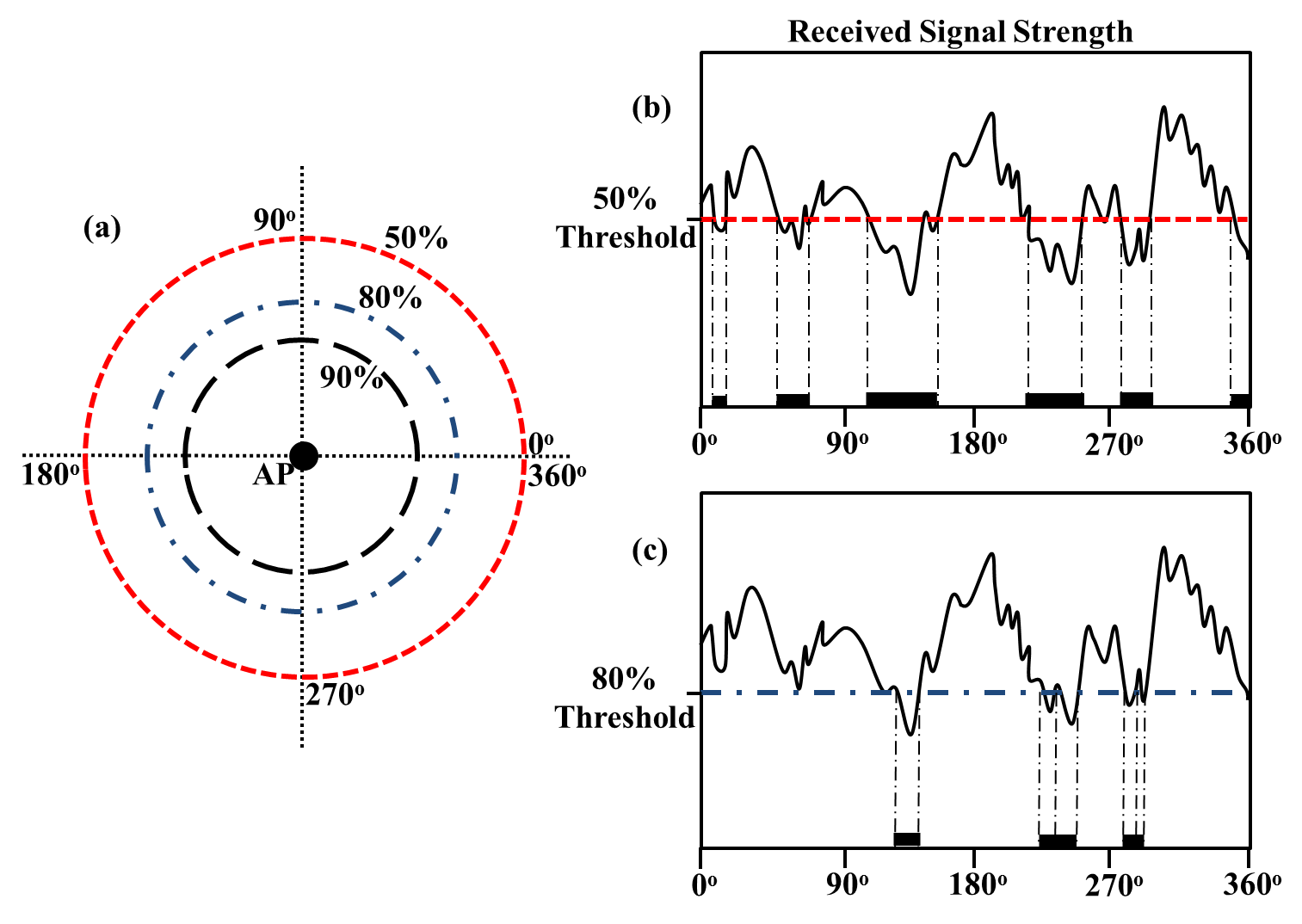} 
 \end{center}
\caption{(a) Coverage regions for 50\%, 80\% and 90\%. (b) Received Signal Strength (RSS) while moving along the 50\% contour. (c) RSS while moving along the 80\% contour. }
\label{fig:1}       
\end{figure*}

This paper considers the effect of the path loss and shadowing. The effect of multipath is neglected and is beyond the scope of this work. Fig. \ref{fig:1} gives a clearer picture of the behavior of wireless signals in a coverage area with coverage probabilities of 90\%, 80\% and 50\%. When an MN moves along contours as shown in Fig. \ref{fig:1} (b) and (c), it may observe poor signal strength at some instants. These slow signal deviations due to shadowing follows a Gaussian distribution and is given as
\begin{equation}
\label{eqn:1}
f(r) = \frac{1}{\sigma\sqrt{2\pi}}\exp\Big[{\frac{-(r-\mu)^{2}}{2\sigma^{2}}}\Big]
\end{equation}
Where ~$\mu$ is the mean value and ~$\sigma^{2}$ is the variance of the Gaussian random variable ~$r$.

Suppose, we have the radius of the WLAN cell which is a continuous random variable R with PDF, ~$f(r)$, as shown in Equation (\ref{eqn:1}) and we desire to evaluate the expectation ~$E[g(R)]$ for some function ~$g(r)$. This entails evaluating the integral,
\begin{equation}
\label{eqn:2}
E[g(R)]=\int_{-\infty}^{\infty} f(r)g(r)\,dr
\end{equation}

Since the integral is not easily tractable by analytical or standard numerical methods, the study approached it by simulating realizations of ~$r_{1}$, ~$r_{2}$, ~$r_{3}$, … ~$r_{n}$ of ~$R$, and since the variance is finite, we apply the law of large numbers to obtain an approximation\cite{RefB3}.

\begin{equation}
\label{eqn:3}
E[g(R)]  \sim \frac{1}{n} \sum_{i=1}^{n} g_{i}(r)
\end{equation}
The expression in Equation (\ref{eqn:3}) gives justification for the Monte-Carlo simulation carried out in the study.
\section{Proposed Scheme}
\label{sec:3}
\begin{figure*}
\begin{center}
 \includegraphics[width=0.9\textwidth]{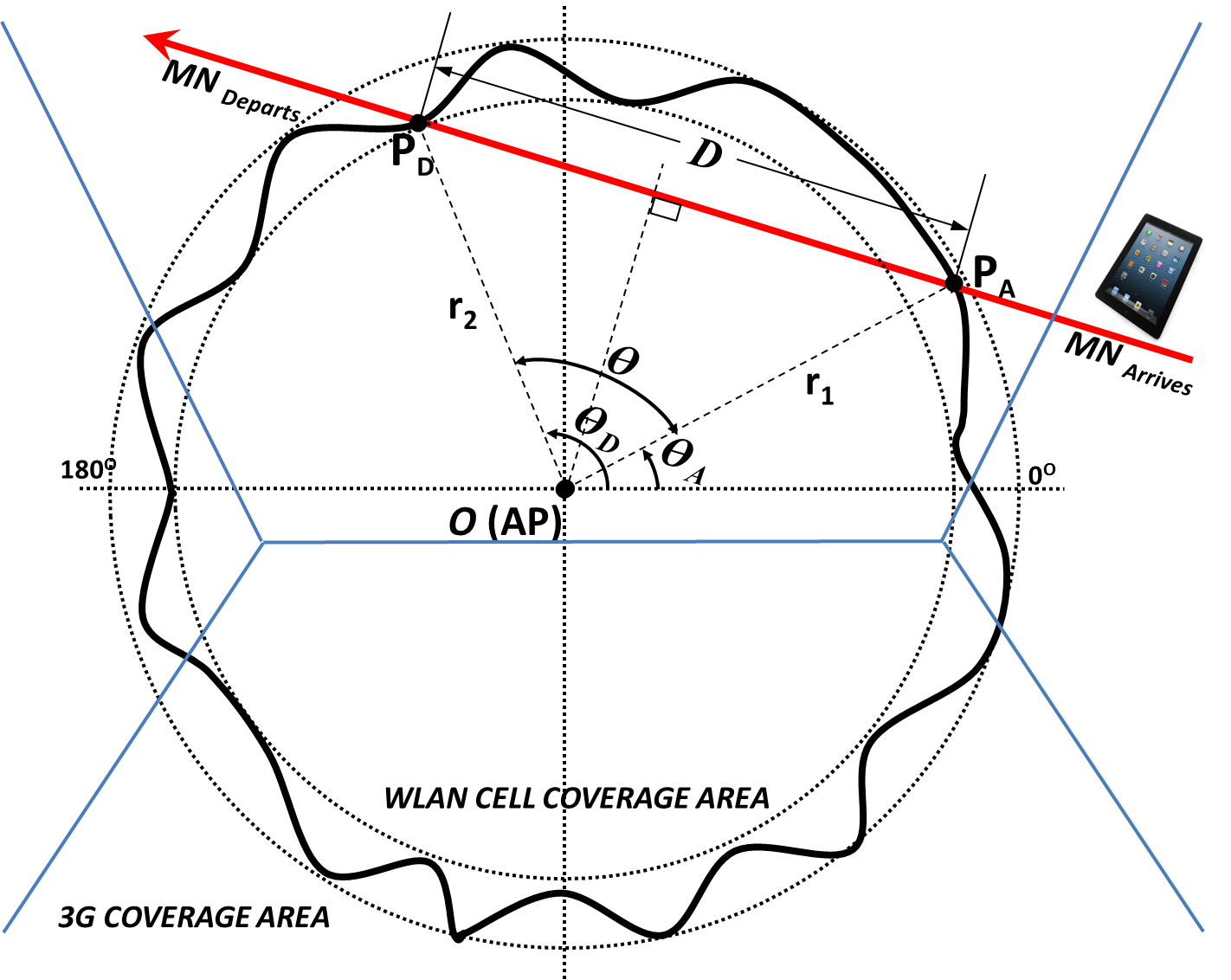}
\caption{A mobile node entering an amoebic WLAN coverage area.}
\label{fig:2}       
\end{center}
\end{figure*}

The work assumes that when an MN is in the coverage area of it's present access network, adjacent to a WLAN cell,\footnote{We present an amoebic based model that extends the ideal circle model employed in previous works \cite{RefJ1,RefJ2,RefJ3,RefJ4}} it may enter the boundary area of the WLAN cell at any point, ~$P_A$ and move along the path ~$|P_AP_D|$, making exit from any point, ~$P_D$ on the coverage boundary (as shown in Fig. \ref{fig:2}). We further assume that the speed,~$v$ of the MN is uniformly distributed in ~$[v_{min}, v_{max}]$. The cell radius is assumed to be stochastic and normally (Gaussian) distributed with defined mean and variance. The justification for having a normal distribution can be given in terms of the Central limit theorem\footnote{The central limit theorem states that given a distribution with mean, ~$\mu$ and variance, ~$\sigma^2$, the sampling distribution of the mean approaches a Gaussian distribution with mean, ~$\mu$ and variance, ~$\frac{\sigma^2}{n}$, where ~$n$ is the number of samples.} \cite{RefB3}, as the total attenuation experienced in a wireless link results from the tallying of several individual shadowing processes forming a Gaussian distribution.

The angle of arrival, ~$\Theta_A$ and angle of departure, ~$\Theta_D$ are assumed to be uniformly distributed within the bound of ~$\Theta$\footnote{We considered ~$[0,\pi]$, while ~$[0,2\pi]$ \cite{RefJ2} and ~$[0,\pi]$ \cite{RefJ3} with angle of arrival with respect to tangential line within ~$[0,\frac{\pi}{2}]$ were considered in previous works.} and we express the angle between random positions of ~$P_A$ and ~$P_D$ as ~$\Theta = | \Theta_D - \Theta_A|$. A realistic coverage area of the WLAN cell with an amoebic structure is considered in this work. As there is only a possibility that the MN moves in and out of the coverage area in any half section of Fig. \ref{fig:2}, we therefore derive an expression to calculate the probability distribution function (PDF) of ~$\Theta$. The PDF of the arrival and departure of the MN from the WLAN coverage at point ~$P_A$ and ~$P_D$ respectively is given by

\begin{equation}
\label{eqn:4}
f_{\Theta_A}(\theta_{A})=\left\{
\begin{array}{c l}
    \frac{1}{\pi}, & 0\leq\Theta_{A}\leq\pi,\\
    0, & Otherwise.
\end{array}\right.
\end{equation}

\begin{equation}
\label{eqn:5}
f_{\Theta_D}(\theta_{D})=\left\{
\begin{array}{c l}
    \frac{1}{\pi}, & 0\leq\Theta_{D}\leq\pi,\\
    0, & Otherwise.
\end{array}\right.
\end{equation}
Since the arrival and departure points of the mobile nodes are independent, the Joint PDF is therefore given as product of their individual marginal functions.

\begin{equation}
\label{eqn:6}
f_{\Theta_A, \Theta_D}(\theta_{A}, \theta_{D})=\left\{
\begin{array}{c l}
    \frac{1}{\pi^{2}}, & 0\leq\Theta_{A}, \Theta_{D}\leq\pi,\\
    0, & Otherwise.
\end{array}\right.
\end{equation}
We find the cumulative distribution function (CDF) of ~$\Theta$ by

\begin{equation}
\label{eqn:7}
\begin{split}
 F_{\Theta}(\theta)&=P(\Theta\leq\theta)\\
    &= \int\int_{\epsilon}f_{\Theta_A, \Theta_D}(\theta_{A}, \theta_{D})\,d\Theta_{D} d\Theta_{A}
\end{split}
\end{equation}

Where ~$\epsilon$ is a set of arrival and departure points along the coverage boundary for the MN such that ~$ 0 \leq \Theta \leq \pi$. ~$P (\Theta\leq\theta) = 0$ for ~$\theta < 0$ and ~$P (\Theta\leq\theta) = 1$ for ~$\theta > \pi$ \cite{RefJ2}. From Fig. \ref{fig:2}, Equation (\ref{eqn:7}) can be expressed as

\begin{equation}
\label{eqn:8}
\begin{split}
 F_{\Theta}(\theta)&=\frac{1}{\pi^{2}}\Bigg(\int_{0}^{\theta}\int_{0}^{\theta + \theta_{D}}d\Theta_{D} d\Theta_{A} + \int_{\theta}^{\pi-\theta}\int_{\theta_{D} - \theta}^{\theta_{D} + \theta}d\Theta_{D} d\Theta_{A}  \\
    &+\int_{\pi - \theta}^{\pi}\int_{\theta_{D} - \theta}^{\pi}d\Theta_{D} d\Theta_{A}\Bigg)
\end{split}
\end{equation}

The final expression of CDF is obtained as:

\begin{equation}
\label{eqn:9}
F_{\Theta}(\theta)= \frac{(2\pi - \theta)\theta}{\pi^{2}}, 0\leq\Theta\leq\pi
\end{equation}

The corresponding PDF of ~$\Theta$ is given by:

\begin{equation}
\label{eqn:10}
f_{\Theta}(\theta)=\left\{
\begin{array}{c l}
    \frac{2(\pi-\theta)}{\pi^{2}}, &   0\leq\Theta\leq\pi,\\
    0, &   Otherwise.
\end{array}\right.
\end{equation}
We can now use the PDF of ~$\Theta$ to compute the PDF of the traversing time by the MN, ~$t_{WLAN}$. Using the Cosine formula, we formulate a geometric expression of the traversing distance, ~$D$ from Fig. \ref{fig:2}
\begin{equation}
\label{eqn:11}
D = \sqrt{r_{1}^{2} +r_{2}^{2} - 2r_{1}r_{2}\cos\theta}
\end{equation}
The traversal distance through the WLAN cell, ~$D$ depends on the traversing angle ~$\theta$.
\begin{equation}
\label{eqn:12}
\begin{split}
 t_{WLAN}&=g(\theta)\\
    &= \frac{\sqrt{r_{1}^{2} +r_{2}^{2} - 2r_{1}r_{2}\cos\theta}}{v}
\end{split}
\end{equation}
Where, $r_{1}$ and $r_{2}$ are the distances of the MN from the access point at the time of entry and exit from the coverage region respectively.

The PDF of the traversing time can thus be expressed as \cite{RefB1}
\begin{equation}
\label{eqn:13}
F(T) = \sum_{i=1}^{n} \biggl|\frac{f(\theta_{i})}{g'(\theta_{i})}\biggl| _{\theta_{i} = g^{-1}(T)}
\end{equation}
Where ~$\theta$ is the root of function ~$g(\theta)$, and ~$g'(\theta)$ is the derivative of ~$g(\theta)$.
\begin{equation}
\label{eqn:14}
\theta = \arccos\Big(\frac{r_{1}^{2} +r_{2}^{2} - t_{WLAN}^{2}v^{2} }{2r_{1}r_{2}}\Big)
\end{equation}
We have the derivative of ~$g(\theta)$ as
\begin{equation}
\label{eqn:15}
g'(\theta) = \frac{r_{1}r_{2}\sin\theta}{v\sqrt{r_{1}^{2} +r_{2}^{2} - 2r_{1}r_{2}\cos\theta}}
\end{equation}
Thus, substituting the Equation (\ref{eqn:14}) into (\ref{eqn:15}) to get,
\begin{equation}
\label{eqn:16}
\begin{split}
 g'(\theta)&=\frac{r_{1}r_{2}\sin\Big[\arccos\Big(\frac{r_{1}^{2} +r_{2}^{2} - t_{WLAN}^{2}v^{2} }{2r_{1}r_{2}}\Big)\Big]}{v\sqrt{r_{1}^{2} +r_{2}^{2} - 2r_{1}r_{2}\cos\Big[\arccos\Big(\frac{r_{1}^{2} +r_{2}^{2} - t_{WLAN}^{2}v^{2} }{2r_{1}r_{2}}\Big)\Big]}}\\
    &= \frac{\sqrt{4r_{1}^{2}r_{2}^{2} - (r_{1}^{2} +r_{2}^{2} - t_{WLAN}^{2}v^{2})^{2}}}{2t_{WLAN}v^{2}}
\end{split}
\end{equation}
To obtain the PDF at ~$\theta$, we substitute Equation (\ref{eqn:14}) into (\ref{eqn:10}),
\begin{equation}
\label{eqn:17}
f(\theta) = \frac{2\Big[\pi-\arccos(\frac{r_{1}^{2} +r_{2}^{2} - t_{WLAN}^{2}v^{2} }{2r_{1}r_{2}})\Big]}{\pi^{2}}
\end{equation}
Thus, from Equations (\ref{eqn:16}) and (\ref{eqn:17}), we can now obtain the PDF of the traversal time, ~$f(T)$, using Equation (\ref{eqn:13}),
\begin{equation}
\label{eqn:18}
f(T) = \frac{4v^{2}t_{WLAN}\Big(\pi-\arccos(\frac{r_{1}^{2} +r_{2}^{2} - t_{WLAN}^{2}v^{2} }{2r_{1}r_{2}})\Big)}{\pi^{2}\sqrt{4r_{1}^{2}r_{2}^{2} - (r_{1}^{2} +r_{2}^{2} - t_{WLAN}^{2}v^{2})^{2}}}
\end{equation}

\section{Handover Decisions}
\label{sec:4}
To have unnecessary handover and handover failure within satisfactory bounds, it is imperative to find two time threshold values, ~$N$ and ~$M$, which correspond to the values for handover decision for unnecessary handover and handover failure respectively. In order to keep the unnecessary handover and handover failure within bounds the handover will only be initiated if the expected traversal time through the WLAN cell exceeds the corresponding threshold value.

\subsection{Probability of Unnecessary Handover}
\label{sec:4a}
This paper attempts to minimize the number of unnecessary handovers. This is achieved by calculating the time threshold value,~$N$, and avoiding handover attempts for which the traversal time through the target network is less than this threshold value. An unnecessary handover is said to occur when the traversing time of an MN in a WLAN cell is smaller than the sum of the handover time into ~$(\tau_A)$ and out of ~$(\tau_D)$ the WLAN coverage area \cite{RefJ3}. We now use the PDF of traversal time obtained in Equation (\ref{eqn:18}) to derive an expression for the CDF of the traversal time, ~$P_u$. This is shown in Equation (\ref{eqn:21}),

\begin{equation}
\label{eqn:19}
P_{u}=\left\{
\begin{array}{c l}
    P_{r}[N<T\leq\tau_{T}], & 0\leq T\leq\frac{(r_{1} +r_{2})}{v},\\
    0, & Otherwise.
\end{array}\right.
\end{equation}

\begin{equation}
\label{eqn:20}
P_{r}[N<T\leq\tau_{T}] = \int^{\tau_{T}}_{N} f(T)\,dt
\end{equation}
Where ~$\tau_{T} = \tau_{A} + \tau_{D}$. The probability of unnecessary handover, ~$P_u$, is expressed as,

\begin{equation}
\label{eqn:21}
\begin{split}
 P_u&=\frac{\Big[2\pi - \arccos(\frac{r_{1}^{2} +r_{2}^{2} - \tau_{T}^{2}v^{2} }{2r_{1}r_{2}})\Big] \arccos\Big(\frac{r_{1}^{2} +r_{2}^{2} - \tau_{T}^{2}v^{2} }{2r_{1}r_{2}}\Big)}{\pi^{2}}  \\
    &-\frac{\Big[2\pi - \arccos(\frac{r_{1}^{2} +r_{2}^{2} - N^{2}v^{2} }{2r_{1}r_{2}})\Big] \arccos\Big(\frac{r_{1}^{2} +r_{2}^{2} - N^{2}v^{2} }{2r_{1}r_{2}}\Big)}{\pi^{2}}
\end{split}
\end{equation}
Let ~$z = \arccos\Big(\frac{r_{1}^{2} +r_{2}^{2} - \tau_{T}^{2}v^{2} }{2r_{1}r_{2}}\Big) $. We obtain the following expression for ~$N$, which is a function of handover latency, velocity, stochastic coverage radius and probability of unnecessary handover.

\begin{equation}
\label{eqn:23}
N = \frac{\sqrt{r_{1}^{2} +r_{2}^{2} - 2r_{1}r_{2}\cos(y)}}{v}
\end{equation}
Where,
\begin{equation}
\label{eqn:22}
y =\pi \pm \sqrt{\pi^{2}(1+P_{u}) -2\pi z + z^{2}}
\end{equation}

We have obtained a new expression for time threshold $N$ for unnecessary handover\footnote{Yan et al.\cite{RefJ2} arrived at a time threshold, ~$t_{WLAN} = \frac{2R}{v}\sin\Big(\arcsin(\frac{v\tau}{2R}-\frac{\pi}{2}P)\Big)$, \\ and Hussain et al.\cite{RefJ3} arrived at, ~$t_{WLAN} = \frac{2Rk}{v\sqrt{1+k^2}}$, where ~$k = \tan\Big[\arctan(\frac{v\tau}{\sqrt{4R^2 - v^2\tau^2}})-\frac{P\pi}{2}\Big]$   }.

\subsection{Probability of Handover Failure}
\label{sec:4b}

Handover failure is said to occur if the handover time into ~$(\tau_A)$ the WLAN cell exceeds the overall time spent by the MN in the WLAN coverage area \cite{RefJ3}. A time threshold, ~$M$, is determined and the probability of handover failure is kept within desirable bounds.

\begin{equation}
\label{eqn:24}
P_{f}  =\left\{
\begin{array}{c l}
    P_{r}[M<T\leq\tau_{A}], & 0\leq T\leq\frac{(r_{1} +r_{2})}{v},\\
    0, & Otherwise.
\end{array}\right.
\end{equation}

\begin{equation}
\label{eqn:25}
P_{r}[M<T\leq\tau_{A}] = \int^{\tau_{A}}_{M} f(T)\,dt
\end{equation}
The probability of handover failure, ~$P_f$, is expressed as,
\begin{equation}
\label{eqn:26}
\begin{split}
 P_f&=\frac{\Big[2\pi - \arccos(\frac{r_{1}^{2} +r_{2}^{2} - \tau_{A}^{2}v^{2} }{2r_{1}r_{2}})\Big] \arccos\Big(\frac{r_{1}^{2} +r_{2}^{2} - \tau_{A}^{2}v^{2} }{2r_{1}r_{2}}\Big)}{\pi^{2}}  \\
    &-\frac{\Big[2\pi - \arccos(\frac{r_{1}^{2} +r_{2}^{2} - M^{2}v^{2} }{2r_{1}r_{2}})\Big] \arccos\Big(\frac{r_{1}^{2} +r_{2}^{2} - M^{2}v^{2} }{2r_{1}r_{2}}\Big)}{\pi^{2}}
\end{split}
\end{equation}

Hence, we have also obtained a new expression for time threshold ~$M$, for handover failure control.

\begin{equation}
\label{eqn:28}
M = \frac{\sqrt{r_{1}^{2} +r_{2}^{2} - 2r_{1}r_{2}\cos(q)}}{v}
\end{equation}

Where,
\begin{equation}
\label{eqn:27}
q =\pi \pm \sqrt{\pi^{2}(1+P_{f}) -2\pi z + z^{2}}
\end{equation}
\begin{figure*}
\begin{center}
 \includegraphics[scale=0.8]{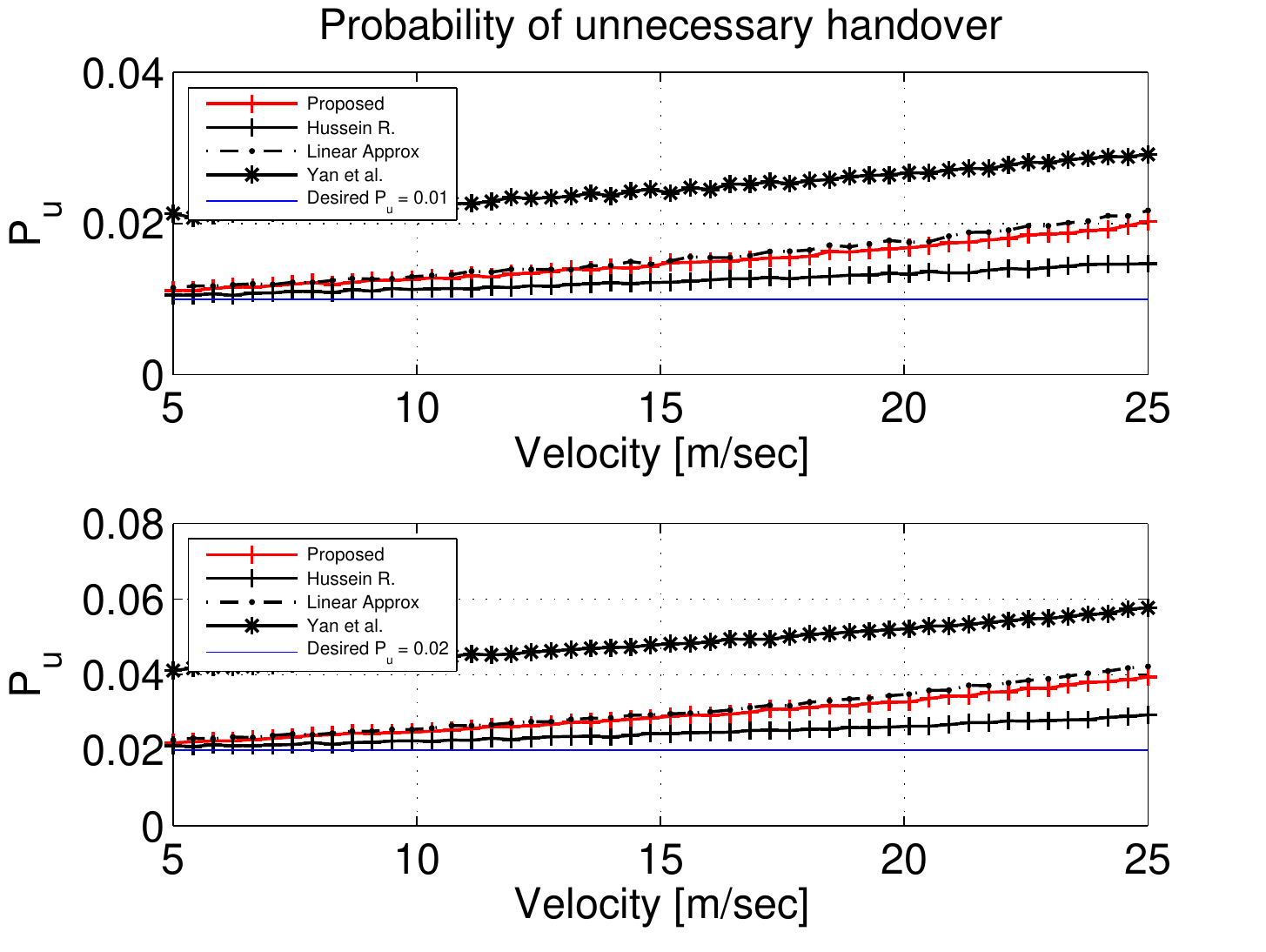}
 \caption{Plot of Probability of Unnecessary Handover vs Velocity of MN.}
\label{fig:3}       
\vspace{0.6cm}
 \includegraphics[scale=0.8]{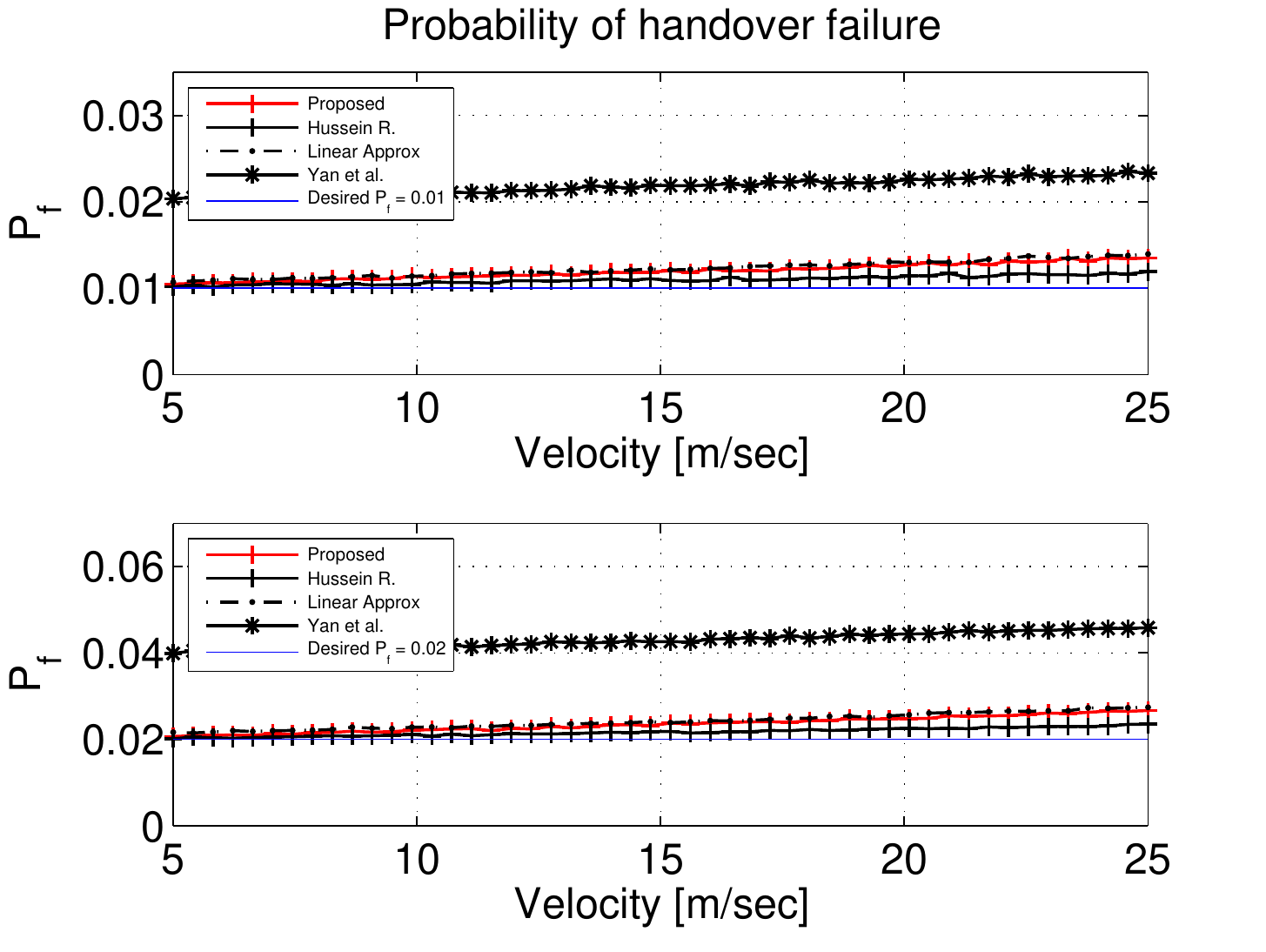}
\caption{Plot of Probability of Handover Failure vs Velocity of MN.}
\label{fig:4}       
\end{center}
\end{figure*}

\section{Results}
\label{sec:5}
To evaluate our proposed model, we performed Monte-Carlo simulations with about 10 million iterations using MATLAB in order to ensure accurate performance. For validation purpose, we also simulated the performance of our proposed model alongside that of other state-of-the-art time prediction vertical handover schemes. Fig. \ref{fig:3} and \ref{fig:4} show the plots of Probability of Unnecessary Handover,~$P_u$ and Probability of Handover Failure,~$P_f$ against the Velocity of MN,~$v$. The threshold values~$(M$ and ~$N)$ obtained in Equations (\ref{eqn:23} and \ref{eqn:28}), the transversal angle,~$\theta$ within ~$[0,\pi]$ bound, dwell time,~$T$ obtained in Equations (\ref{eqn:12}) were used in the experiments. From the graphs, we observe that as the velocity of the MN increases, the probability of unnecessary handover and handover failure increases and deviates from the designed level. This implies that speed has an impact on the prediction of threshold values, which are obtained using this probabilistic model. This deviation in probabilities of handover failure and unnecessary handover are in compliance with the earlier existing works~\cite{RefJ1,RefJ2,RefJ3,RefJ4}.

Results show that the proposed model performs closely to the Linear approximation method employed in \cite{RefJ1} which considered a circular coverage cell. Our work out-performed results of Yan et al \cite{RefJ2}, which considered a ~$[0,2\pi]$ bound, but under-performed when compared to the work of Hussain \emph{et al.} \cite{RefJ3}, which also considered the angle of arrival and departure to lie between ~$[0,\pi]$. However, this model\cite{RefJ3} is not only unrealistic but also impractical as it requires precise information (on the tangential angle of arrival of the MN which is uniformly distributed between ~$[0,\frac{\pi}{2}]$) from the system. Despite the performance limitations, our work considered the effect of slow fading and presents a realistic depiction of the WLAN coverage area with the view that the wireless environment is a stochastic one with numerous uncertainties, the proposed model gives a more accurate prediction than previous works in literature.

%

\section{Conclusion and Future work}
\label{sec:6}
This paper has presented new models for realistic renderings of the WLAN coverage area. The geometric-based model for HNE in this paper combines and extends theoretical results from previous mathematical analysis conducted by several researchers. The resulting model is probabilistic and based on various network parameters which include the random varying cell radius, the traverse angle, ~$\theta$, the velocity of the MN. This model is also unique in the sense that it can simulate different coverage scenarios with respect to dwell time of the MN in the WLAN cell. As all parameters of the models were derived from extensive geometric and probability analysis, they correctly simulate the actual behavior of the MN traversing a WLAN coverage area. From results obtained, we arrive at the conclusion that slow fading (shadowing) has minimal effect on vertical handover models. Our future work will consider the effects of fast fading due to multi-path.

The models presented were validated by comparing simulated results with works of other researchers under similar conditions. The quality of these simulations qualitatively matched the actual behaviors of MN traversing a realistic WLAN cell. To the best of our knowledge, this is the first geometric-based model that considers an amoebic cell structure, i.e. an irregular shaped with probabilistic coverage boundary, for simulating the probability of handover. The model was successful in minimizing Probability of Unnecessary Handover,~$P_u$ and Probability of Handover Failure,~$P_f$. These geometric-based models for HNE are also the first model of its kind in the existing literatures.


\end{document}